\documentstyle[eqsecnum,twocolumn,aps]{revtex}
\begin{document}
\draft
%\preprint {}
\title {Nonequilibrium phase transition in the kinetic Ising model:
Divergences\\ 
of fluctuations and responses near the transition point}
\author {Muktish Acharyya$^*$}
\address {Department of Physics\\
Indian Institute of Science, Bangalore-560012, India\\
and\\
Condensed Matter Theory Unit\\
Jawaharlal Nehru Centre for Advanced Scientific Research\\
Jakkur, Bangalore-560064, India}

\date{\today}
\maketitle
\begin{abstract}
The nonequilibrium dynamic phase transition, in the kinetic Ising model 
in presence of an oscillating magnetic field,
has been studied by Monte Carlo simulation. 
The fluctuation of dynamic order 
parameter has been studied as a function
of temperature near the dynamic transition point. The temperature variation
of appropriately defined 'susceptibility' has also been studied near the
dynamic transition point. Similarly, fluctuation of energy and appropriately
defined 'specific-heat' have been studied as a function of temperature near
the dynamic transition point. In both the cases, the fluctuations (of dynamic
order parameter and energy) 
and the corresponding responses diverge
(as power law fashion) near the dynamic
transition point with similar critical behavior
(with identical exponent values).
\end{abstract}

\vspace {0.4 cm}

\narrowtext
\section{Introduction}
The physics of equilibrium phase transition in the Ising model is
well understood \cite{st}. However, the mechanism behind the
nonequilibrium phase transition is not yet explored rigorously and the basic
phenomenology is still undeveloped. It is quite interesting to study
how the system behaves if it is driven out of equilibrium. 
The simplest prototype example is the kinetic Ising model in oscillating
magnetic field.
In this context,
the dynamic response of the Ising system in presence of an oscillating 
magnetic field has been studied
extensively \cite{rkp,dd,smp,tom,lo,ac} in the last few years. 
The dynamic hysteresis 
\cite{rkp,dd,smp} and the nonequilibrium
dynamic phase transition \cite{tom,lo,ac} are two important aspects of
the dynamic response of the kinetic Ising model in presence of an
oscillating magnetic field.  
The nonequilibrium dynamic phase transition in the kinetic Ising model
in presence of an oscillating magnetic field, was first studied by
Tome and Oliviera \cite{tom}.
They solved the mean-field (MF)
dynamic equation of motion (for the average magnetization) of the kinetic
Ising model in presence of a sinusoidally oscillating magnetic field.
By defining the dynamic order parameter as the time averaged
magnetization over a full cycle of the 
oscillating magnetic field, they showed that depending upon
the value of the field amplitude and the temperature, the 
dynamic order parameter takes
nonzero value from a zero value. 
In the field amplitude and temperature plane there
exists a distinct phase boundary separating dynamic ordered 
(nonzero value of order parameter) and disordered (order 
parameter vanishes) phase.  
A tricritical point (TCP),
(separating the nature (discontinuous-continuous) of the transition)
on the phase boundary line,
was also observed by them \cite{tom}. However, 
one may argue that such a mean-field transition
is not truly dynamic in origin since it exists even in the 
quasi-static (or zero
frequency) limit. This is because, if the field amplitude is less than the
coercive field (at temperature less than the transition temperature
without any field), then the response magnetization varies periodically
but asymmetrically even in the zero frequency limit; the system remains locked
to one well of the free energy and cannot go to the other one, in the absence
of fluctuation.

The true dynamic nature of this kind of phase transition (in presence of 
fluctuation) was first attemted 
to study by Lo and Pelcovits \cite{lo}. They have studied
the dynamic phase transition in the kinetic Ising model in presence of an
oscillating magnetic field by Monte Carlo (MC) simulation which
allows the microscopic fluctuations.
Here, the transition disappears in the 
zero frequency limit; due to the fluctuations, the magnetization flips to the
direction of the magnetic field and the dynamic order parameter (time
averaged magnetization) vanishes.
However, they \cite{lo} have 
not reported any precise phase boundary.
Acharyya and Chakrabarti \cite{ac} studied the nonequilibrium dynamic phase
transition in the kinetic Ising model 
in presence of oscillating magnetic field by 
extensive MC simulation. 
They \cite{ac} have also identified 
that this dynamic phase transition (at a particular
nonzero frequency of the oscillating magnetic field) is associated with the 
breaking of the symmetry of
the dynamic hysteresis ($m-h$) loop. 
In the dynamically disordered phase 
(where the value of order parameter vanishes)
the corresponding hysteresis loop is
symmetric, and loses its symmetry in the ordered phase (giving
nonzero value of dynamic order parameter).
They \cite{ac} also studied the temperature variation of the ac susceptibility
components near the dynamic transition point.  
The major observation was
that the imaginary (real) part of the ac susceptibility gives a
peak (dip) near the dynamic transition point (where the dynamic
order parameter vanishes). The 
important conclusions were: (i) this is a distinct
signal of phase transition and (ii) this is an indication of the
thermodynamic nature of the phase transition.
The Debye relaxation of the dynamic order parameter 
and the critical slowing down have been studied very recently
\cite{ma} both by MC simulation and by solving the dynamic MF equation 
\cite{tom} of
motion for the average magnetization.
The specific-heat singularity \cite{ma} 
near the dynamic transition point is also
an indication of the thermodynamic nature of this dynamic phase transition.
It is worth mentioning here that the statistical distribution of dynamic
order parameter has been studied by Sides et al \cite{rik}. The nature of the
distribution changes
(from bimodal to unimodal) near the dynamic transition point. They have
also observed 
\cite{rik} that the fluctuation of 
the hysteresis loop area becomes considerably large near the dynamic
transition point.

In the case of equilibrium phase transitions, the fluctuation - dissipation
theorem (FDT) states 
(due to the applicability of Gibbs formalism) that the 
mean square fluctuations
of some intrinsic physical quantities (say, energy, magnetization etc.) are
directly related with some 
responses (specific heat, susceptibility
etc.) of the system. Consequently, near the 
ferro-para transition point, both the fluctuation of
magnetization and the susceptibilty show same singular behavior. If it
is of power law type, the same singular behavior will be characterised by
the same exponent. This is also true for fluctuation of energy and 
the specific heat. These are the consequences of fluctuation-dissipation
theorem \cite{st}. Here, the main motivation is to study the fluctuations
and corresponding responses near the dynamic transition temperature.

In this paper, the fluctuations of dynamic order parameter and the energy
are studied as a function of temperature near the dynamic transition point.
The temperature variations of `susceptibility' and the 'specific-heat' are
also studied near the transition point. 
The temperature variation of the fluctuation of dynamic order
parameter and 
that of the `susceptibility' are compared. Similarly, the 
temperature variation of the fluctuation
of energy and 
that of the `specific-heat' are compared. The paper is organised as
follows: the model and the simulation scheme are discussed in section II,
the results are reported in Section III, section IV contains the summary
of the work.

\section{Model and simulation}

\noindent The Ising model with nearest neighbor ferromagnetic coupling 
in presence
of a time varying magnetic field can be represented by the Hamiltonian

\begin{equation}
H = -\sum_{<ij>} J_{ij} s_i^z s_j^z - h(t) \sum_i s_i^z
\label{hm} 
\end{equation}

Here, $s_i^z (=\pm 1)$ is Ising spin variable, $J_{ij}$ is the interaction
strength and $h(t) = h_0 {\rm cos}(\omega t)$  
represents the oscillating magnetic field, where
$h_0$ and $\omega$ are the amplitude and the frequency 
respectively of the oscillating field. The system
is in contact with an isothermal heat bath at temperature $T$. For simplicity
all $J_{ij} (= J > 0)$ are taken equal to 
unity and the boundary condition is chosen to be periodic. The temperature 
($T$) is measured in the unit of $J/K$, where $K$ is the Boltzmann constant
(here $K$ is taken unity).

A square lattice of linear size $L (=100)$ has been considered.
At any finite
temperature $T$ and for a fixed frequency ($\omega$) 
and amplitude ($h_0$) of the
field, the microscopic
dynamics of this system has been studied here by Monte Carlo
simulation using Glauber single spin-flip dynamics
with a particular choice of the Metropolis rate of single spin-flip \cite{mc}. 
Starting from an initial condition where all spins are up, each lattice site is
updated here sequentially and one such full scan over the entire lattice is
defined as the unit time step (Monte Carlo step or MCS).
The instanteneous magnetization
(per site),
 $m(t) = (1/L^2) \sum_i s_i^z$ has been calculated. From the instanteneous
magnetization, the dynamic order parameter $Q = {\omega \over {2\pi}}
\oint m(t) dt$ (time averaged magnetization over a full cycle of the
oscillating field) is calculated. 
Some of the transient loops have been discarded
to get the stable value of the dynamical quantities.

\section{Results}

\subsection{Temperature variations of susceptibility and
fluctuation of dynamic order 
parameter }

The fluctuation of the dynamic order parameter is
$$\delta Q^2 = \left(<Q^2> - <Q>^2\right),$$
\noindent where the $< >$ stands for the averaging over various Monte Carlo
samples.

The `susceptibility' is defined as
$$\chi = -{d<Q> \over dh_0}.$$

Here, a square lattice of linear size $L$ (=100) has been
considered. $<Q^2>$ and $<Q>$ are calculated using MC simulation.
The averaging has been done over 100 different (uncorrelated) MC
samples.

The temperature variations of fluctuation of $Q$, i.e.,
$\delta Q^2$ and `susceptibility' $\chi$ have
been studied here and both plotted in Fig. 1. From the figure
it is observed that both $\delta Q^2$ and $\chi$ 
diverge near the dynamic
transition point (where $Q$ vanishes). 

This has been studied for two different values of field amplitude $h_0$
(Fig. 1a is for $h_0$ = 0.2 and Fig. 1b is for $h_0$ =0.1). 
The dynamic transition temperatures $T_d(h_0)$, at which $\chi$ and
$\delta Q^2$ diverge, are 1.91$\pm0.01$ for $h_0$ = 0.2 and $2.15\pm0.01$
for $h_0$ = 0.1. These values of $T_d(h_0)$ agree with the phase
diagram estimated from vanishing of $Q$.
The $log_e (\chi)$ versus $log_e (T_d - T)$ and
$log_e (\delta Q^2)$ versus $log_e (T_d - T)$ 
plots show (insets of Fig. 1) that
$\chi \sim (T_d-T)^{-\alpha}$ and $\delta Q^2 \sim (T_d-T)^{-\alpha}$. 
For $h_0$ = 0.2, $\alpha \sim 0.53$ 
(inset of Fig. 1a) and
for $h_0$ = 0.1, $\alpha \sim 2.5$  
(inset of Fig. 1b). 
Results show that both $\chi$ and $\delta Q^2$
diverge near $T_d$ as a power law with the same exponent $\alpha$, 
though there is a crossover region (where the effective exponent values are
different).

\subsection{Temperature variations of specific-heat and
fluctuation of energy}

 The time averaged (over a full cycle) cooperative energy of the system
is

$$E = -(\omega/{2 {\pi} L^2}) \oint 
\left(\sum_{i,j} s_i^z s_j^z \right) dt,$$ 
\noindent and the fluctuation of the cooperative energy is
$$\delta E^2 =\left(<E^2> - <E>^2\right).$$

\noindent The `specific-heat' $C$ \cite{ma} is defined as the derivative of
the energy (defined above) with respect to the temperature, and
$$C = {d<E> \over dT}.$$

Here, also a square lattice of linear size $L$ (=100) has been
considered. $<E^2>$ and $<E>$ are calculated using MC simulation.
The averaging has been done over 100 different (uncorrelated) MC
samples.

The temperature variation of the 'specific-heat' has been studied \cite{ma}
and found prominent divergent behavior near the dynamic transition point
(where $<Q>$ vanishes). The temperature variation of fluctuation of energy,
$(\delta E)^2$ has been studied and plotted in Fig. 2. From the figure
it is clear that the mean square fluctuation of energy ($\delta E^2$) and
the specific heat ($C$) both diverge near the dynamic transition point
(where the dynamic order parameter $Q$ vanishes)

This has been studied for two different values of field amplitude $h_0$
(Fig. 2a is for $h_0$ = 0.2 and Fig. 2b is for $h_0$ =0.1). 
Here also (like the earlier case) the specific heat $C$ and $\delta E^2$
are observed to diverge at temperatures $T_d$. 
The temperatures $T_d(h_0)$, at which $C$ and $\delta E^2$ diverge, are
1.91$\pm0.01$ for $h_0$ = 0.2 (Fig. 2a)
and 2.15$\pm$0.01 for $h_0$ = 0.1 (Fig. 2b).
These values also agree with the phase diagram estimated from vanishing
of $Q$.
The $log_e (C)$ vs. $log_e (T_d - T)$ and
$log_e (\delta E^2)$ vs. $log_e (T_d - T)$ plots show (insets of Fig. 2) that
$C \sim (T_d-T)^{-\gamma}$ and $\delta E^2 \sim (T_d-T)^{-\gamma}$. 
For $h_0$ = 0.2, $\gamma \sim 0.35$  
(inset of Fig. 2a) and
for $h_0$ = 0.1, $\gamma \sim 0.43$  
(inset of Fig. 2b).
Like the earlier case, here 
also the results show that both $\chi$ and $\delta Q^2$
diverge near $T_d$ as a power law with the same exponent $\alpha$, 
though there is a crossover region (where the effective exponent values are
different).

\bigskip
\section{Summary}

The nonequlibrium dynamic phase transition, in the kinetic Ising model
in presence of an oscillating magnetic field, is studied by Monte
Carlo simulation.
Acharyya and Chakrabarti \cite{ac} observed that the complex susceptibility
components have peaks (or dips) at the dynamic transition point. Sides et
al \cite{rik} observed that the fluctuation in the hysteresis loop area grows
(seems to diverge) near the dynamic transition point.
It has been observed \cite{ma} that the 'relaxation time' and the appropriately
defined 'specific-heat' diverge near the dynamic transition
point. 

The mean square fluctuation of dynamic order parameter 
and the `susceptibility' are
studied as a function of temperature, near the dynamic transition point. 
Both shows the power law variation with respect to the reduced temperature
near the dynamic transition point with 
the same exponent values.
Similar,
observation has been made for the case of 
mean square fluctuation of energy and the
`specific-heat'. 
It appears that although the effective exponent values for the fluctuation
and the appropriate linear resoponse differ considerably, away from
the dynamic transition point $T_d$, they eventually converge and give 
identical value as the temperature interval $|T_d-T|$ decreases and
falls within a narrow crossover region.
These numerical observations indicate that the fluctuation-
dissipation relation \cite{st}
holds good in this case of the nonequilibrium
phase transition in the kinetic Ising model. 
However, at this stage there is no analytic support of the FDT 
in this case.

Finally, it should be mentioned, in this context, 
that experiments \cite{rpi1} on ultrathin
ferromagnetic Fe/Au(001) films have 
been performed to study the frequency dependence
of hysteresis loop areas.
Recently, attempts have been made \cite{rpi2} to measure
the dynamic order parameter $Q$ experimentally, 
in the same material, by extending their previous study \cite{rpi1}.
The dynamic phase transition 
has been studied from the observed temperature variation of $Q$.
However, the detailed investigation
 of the dynamic phase transitions by measuring
variations of associated response functions (like the ac susceptibility,
specific-heat, correlations, relaxations etc) have not yet been performed
experimentally.

\vspace {0.3 cm}
\section*{Acknowledgments}

The author would like to thank B. K. Chakrabarti for critical remarks and
for a careful reading of the manuscript. 
The JNCASR, Bangalore, is gratefully acknowledged for financial support and
for computational facilities.

\vskip 2 cm

\centerline {\large {\bf Figure Captions}}

\vskip 1 cm

\noindent Fig.1 Temperature variations of $<Q>$, $\chi$ and
$\delta Q^2$ 
for two different values of field amplitudes $h_0$:
(a) For $h_0$ = 0.2; $Q$ (solid line), $\chi$ (circles) and
$\delta Q^2$ (triangles). (b) For $h_0$ = 0.1; $Q$ (solid line),
$\chi$ (circles) and $\delta Q^2$ (triangles).
All the data points are plotted in arbitrary units.
Here, $\omega$ = $2 \pi \times$ 0.01.
Corresponding insets
show the plots of  $log_e (\chi)$ (circles) and
$log_e (\delta Q^2)$ (triangles)
against $log_e (T_d-T)$. Solid lines represent the linear best-fit
in the region very closed to $T_d$. 

\vskip 1.0 cm

\noindent Fig.2 Temperature variations of $<Q>$, $C$ and 
$\delta E^2$ 
for two different values of field amplitudes $h_0$:
(a) For $h_0$ = 0.2; $Q$ (solid line), $C$ (circles) and
$\delta E^2$ (triangles). (b) For $h_0$ = 0.1; $Q$ (solid line),
$C$ (circles) and $\delta E^2$ (triangles).
All the data points are plotted in arbitrary units.
Here, $\omega$ = $2 \pi \times$ 0.01.
Corresponding insets
show the plots of $log_e (C)$ (circles) and
$log_e (\delta E^2)$ (triangles)
against $log_e (T_d-T)$. Solid lines represent the linear best-fit
in the region very closed to $T_d$.
\end{document}